\documentclass[conference]{IEEEtran}
\IEEEoverridecommandlockouts

\usepackage{amsmath}
\usepackage{amsfonts}
\usepackage{amssymb}
\usepackage{algorithm}
\usepackage{algorithmic}
\usepackage{verbatim}
\usepackage{hyperref}
\usepackage{multirow}
\usepackage{framed}

\usepackage{graphicx}
\usepackage{epstopdf}
\graphicspath{{fig/}}
\usepackage{calc}
\usepackage{color}
\usepackage[caption=false,font=footnotesize]{subfig}

\newtheorem{theorem}{Theorem}

\newtheorem{example}{Example}

\newtheorem{stopping criterion}{stopping criterion}

\newcommand{\beq}{\begin{equation}}
\newcommand{\eeq}{\end{equation}}
\newcommand{\beqnn}{\begin{equation*}}
\newcommand{\eeqnn}{\end{equation*}}
\newcommand{\beqy}{\begin{eqnarray}}
\newcommand{\eeqy}{\end{eqnarray}}
\newcommand{\beqynn}{\begin{eqnarray*}}
\newcommand{\eeqynn}{\end{eqnarray*}}
\newcommand{\bit}{\begin{itemize}}
\newcommand{\eit}{\end{itemize}}
\newcommand{\ben}{\begin{enumerate}}
\newcommand{\een}{\end{enumerate}}
\newcommand{\bex}{\begin{example}}
\newcommand{\eex}{\end{example}}

\newcommand{\balg}[1]{\begin{algorithm} \caption{#1}}
\newcommand{\ealg}{\end{algorithm}}

\newcommand{\balgc}{\begin{algorithmic}[1]}
\newcommand{\ealgc}{\end{algorithmic}}

\newcommand{\bary}{\begin{array}}
\newcommand{\eary}{\end{array}}
\newcommand{\bmx}{\begin{bmatrix}}
\newcommand{\emx}{\end{bmatrix}}
\newcommand{\bsmx}{\left[\begin{smallmatrix}}
\newcommand{\esmx}{\end{smallmatrix}\right]}
\newcommand{\bmxc}[1]{\left[\begin{array}{@{}#1@{}}}
\newcommand{\emxc}{\end{array}\right]}
\newcommand{\bcn}{\begin{center}}
\newcommand{\ecn}{\end{center}}

\newcommand{\Rbb}{{\mathbb{R}}}

\newcommand{\bigO}{{\mathcal{O}}}

\newcommand{\G}{\boldsymbol{G}}

\newcommand{\I}{\boldsymbol{I}}

\newcommand{\R}{\boldsymbol{R}}

\renewcommand{\a}{\boldsymbol{a}}

\newcommand{\e}{\boldsymbol{e}}

\newcommand{\h}{\boldsymbol{h}}

\newcommand{\n}{\boldsymbol{n}}
\newcommand{\p}{\boldsymbol{p}}

\newcommand{\bst}{\boldsymbol{t}} 

\renewcommand{\v}{\boldsymbol{v}}

\newcommand{\0}{{\boldsymbol{0}}}

\newcommand{\norm}[1]{\left\lVert #1 \right\rVert} 

\usepackage{slashbox}

\begin{document}

\title{A Linearithmic Time Algorithm for a Shortest Vector Problem in Compute-and-Forward Design}

\author{\IEEEauthorblockN{Jinming~Wen\IEEEauthorrefmark{1} and
Xiao-Wen~Chang\IEEEauthorrefmark{2} }
\IEEEauthorblockA{\IEEEauthorrefmark{1}ENS de Lyon, LIP, (CNRS, ENS de Lyon, Inria, UCBL),
Lyon 69007, France, Email: jinming.wen@ens-lyon.fr}
\IEEEauthorblockA{\IEEEauthorrefmark{2}School of Computer Science, McGill University,
Montreal, H3A 2A7, Canada, E-mail: chang@cs.mcgill.ca}
\thanks{This work was supported by NSERC of Canada grant 217191-12, "Programme Avenir
Lyon Saint-Etienne de l¡¯Universit\'e de Lyon" in the framework of the programme
"Inverstissements d'Avenir" (ANR-11-IDEX-0007) and by ANR through the HPAC project under
Grant ANR~11~BS02~013. }}


%


\maketitle

\begin{abstract}
We propose an algorithm with expected complexity of $\bigO(n\log n)$ arithmetic operations to solve a special shortest vector problem
arising in computer-and-forward design, where $n$ is the dimension of the channel vector.
This algorithm is  more efficient than the best known algorithms with proved complexity.

\end{abstract}


%
\IEEEpeerreviewmaketitle

\section{Introduction}

In this paper, we consider solving the following quadratic integer programming problem
arising in compute-and-forward (CF) design:
\begin{align}
\label{e:CFP}
\min_{\a \in\mathbb{Z}^n\backslash\{\0\}} \a^T\G\a,
\mbox{ where }\G =\I -\frac{P}{ 1+P\|\h\|^2 }\h\h^T,
\end{align}
where $P$, a constant, is the transmission power,
$\h\in \mathbb{R}^n$ is a random channel vector following the normal distribution $\mathcal{N}(\0, \I )$ and $\|\h\| = (\h^T\h)^{1/2}$.

In relay networks, CF is a promising relaying strategy that can offer higher rates than traditional ones
such as amplify-and-forward and decode-and-forward, especially in the moderate SNR regime.
To find the optimal coefficient vector that maximizes
the computation rate at a relay in CF scheme, we need to solve \eqref{e:CFP}.

It is easy to verify that the matrix $\G$ in \eqref{e:CFP} is symmetric positive definite,
so it has the Cholesky factorization $\G = \R^T\R$, where $\R$ is an upper triangular matrix.
Then we can rewrite \eqref{e:CFP} as a shortest vector problem (SVP):
\begin{equation}
\label{e:SVP}
\min_{\a \in\mathbb{Z}^n\backslash\{\0\}}\| \R \a\|^2.
\end{equation}

The general SVP  arises in  many applications, including cryptography and communications,
and there are different algorithms to solve it
(see, e.g.,  \cite{AgrEVZ02, HanPS11}).
Although it has not been proved that the general SVP is NP-hard,
it was shown in \cite{Ajt98} that  the SVP is NP-hard for randomized reductions.
However, since the SVP \eqref{e:CFP} is special due to the structure of $\G$,
efficient algorithms can be designed to solve it.

Various methods have been proposed for solving \eqref{e:CFP}, including
the  branch-and-bound algorithm \cite{RicSJ12} (which did not use the properties of $\G$),
the algorithm proposed in \cite{SahG14} and its improvement \cite{SahG15},
which has the best known proved expected complexity of $\bigO(n^{1.5} \log n)$
(In this paper, the complexity is measured by the number of arithmetic operations),
the sphere decoding based algorithm given in \cite{WenZMC15}, whose expected complexity is
approximately  $\bigO(n^{1.5})$.
There are also some suboptimal algorithms, see \cite{SakVHB12}, \cite{ZhoM14}  and \cite{ZhoWM14}.

In this paper, we will modify the algorithm proposed in \cite{SahG15}
for solving \eqref{e:CFP} to reduce the expected complexity to
$\bigO(n\log n)$.

The rest of the paper is organized as follows.
In Section \ref{sec:EA}, we review the algorithms proposed in \cite{SahG15}  and \cite{WenZMC15}
for solving \eqref{e:CFP}.
Then, in Section \ref{sec:PA}, we propose a new algorithm.
To compare these three algorithms computationally, we give numerical results  in Section \ref{sec:sim}.
Finally, conclusions are given in Section \ref{sec:Con}.

{\it Notation.}
We use $\mathbb{R}^n$ and $\mathbb{Z}^n$ to denote the spaces of the $n-$dimensional column real vectors and integer vectors, respectively,
and $\mathbb{R}^{m\times n}$ to denote the spaces of the $m\times n$ real matrices, respectively.
Boldface lowercase letters denote column vectors and boldface uppercase letters denote matrices.
For a vector $\bst$, $\norm{\bst}$ denotes its $\ell^2$-norm.
We use $\e_k$ to denote the $k$-th column of an $n\times n$ identity matrix $\I$,
$\e$ to denote the $n-$dimensional vector with all of its elements being one
and $\0$ denote the $n-$dimensional zero vector.
For $x\in \mathbb{R}$, we use $\lceil x\rceil$ and $\lfloor x\rfloor $ respectively denote the smallest integer that is not smaller than $x$
and the largest integer that is not larger than $x$.
For $\bst\in \mathbb{R}^n$, we use $\lfloor \bst\rceil$ to denote its nearest integer vector, i.e.,
each entry of $\bst$ is rounded to its nearest integer (if there is a tie, the one with smaller magnitude is chosen).
For a set $\Phi$, we use $|\Phi|$ denote its cardinality.

\section{Existing Algorithms}
\label{sec:EA}
In this section, we review the algorithms proposed in \cite{SahG15}  and \cite{WenZMC15}
for solving \eqref{e:CFP}.

\subsection{The Algorithm of of Sahraei and Gastpar}
\label{sec:EA1}
In this subsection, we review the algorithm of  Sahraei and Gastpar proposed in \cite{SahG15},
which has the complexity $\bigO(n\psi\log(n\psi))$,
where
\beq
\label{e:psi}
\psi=\sqrt{1+P\|\h\|^2}.
\eeq
We will show that the expected complexity is  $\bigO(n^{1.5}\log n)$
when $\h$, as it is assumed in computer-and-forward design, follows the normal distribution $\mathcal{N}(\0, \I )$.
For the sake of convenience, we will refer to the algorithm as Algorithm SG.

Given $\h$, we assume it is nonzero. Otherwise the problem \eqref{e:CFP} becomes trivial.
We first simplify notation as \cite{SahG15} does. Let
\beq
\label{e:valpha}
\v=\frac{1}{\|\h\|}\h, \quad \alpha=\frac{P \|\h\|^2 }{1+P \|\h\|^2 }.
\eeq
Then $\|\v\|=1$ and we can rewrite \eqref{e:CFP} as
\beq
\label{e:oldop}
\min_{\a\in\mathbb{Z}^n\backslash\{\0\}}\a^T\G\a, \mbox{ where }
\G = \I - \alpha\v \v^T.
\eeq

The algorithm in \cite{SahG15} is based on the following two theorems which were given in \cite{NazG11} and \cite{SahG14}, respectively.
\begin{theorem}
\label{t:abound}
The solution $\a^\star$ to the SVP problem \eqref{e:oldop} satisfies
$$
\|\a^\star\|\leq \psi,
$$
where $\psi$ is defined in \eqref{e:psi}.
\end{theorem}

\begin{theorem}
\label{t:solpr}
The solution $\a^\star$ to \eqref{e:oldop} is a standard unit vector, up to a sign, i.e., $\a^\ast =\pm \e_i$
for some $i \in \{1,\ldots,n\}$, or satisfies
\beq
\label{e:xbound}
\a^\star-\frac{1}{2}\e<\v x<\a^\star+\frac{1}{2}\e
\eeq
for some $x \in \Rbb$, leading to
\beq
\label{e:abound}
\a^\star=\lceil \v x\rfloor.
\eeq
\end{theorem}


Define $\a(x)=\lceil \v x\rfloor$, then for any $v_i\neq 0$,
$a_i(x)=\lceil v_i x\rfloor$ is a piecewise constant function of $x$ and so is the objective function
$f(\a(x))=\a(x)^T\G\a(x)$.
Thus, $f(\a(x))$ can be represented as
\begin{equation}
\label{e:fax}
f(\a(x))=
\begin{cases}
p_k, &\mbox{if $\xi_k<x<\xi_{k+1}, \,\;k=0,1,...$}\\
q_k, &\mbox{if $x=\xi_k, \,\;k=0,1,...$}
\end{cases} ,
\end{equation}
where $\xi_k$ are sorted real numbers denoting the discontinuity points of $f(\a(x))$.
By \eqref{e:fax}, $f(\a(x))$ is a constant for $x \in (\xi_k, \xi_{k+1})$.
Thus, by \eqref{e:xbound},
\beq
\label{e:fastar}
f(\a^\star)=\min_{k=0,1,...} f\left(\a \Big(\frac{\xi_k+\xi_{k+1}}{2}\Big)\right).
\eeq

To reduce the computational cost,  \cite{SahG15} looks at only part of the discontinuity points.
It is easy to see that the discontinuity points of $a_i(x)$ are $x=\frac{c}{|v_i|}$ (where $v_i \neq 0$)
for any $c-\frac{1}{2} \in \mathbb{Z}$, which are also the discontinuity points of the objective function $f(\a(x))$.
Notice that if $\a^\ast$ is a solution, then $-\a^\ast$ is also a solution.
(This fact was used in  \cite{WenZMC14} to reduce the search cost.)
Using this fact, \cite{SahG15} just considers only positive discontinuity points,
i.e., only positive candidates for $c$ are  considered.
Furthermore, from Theorem \ref{t:abound},
$$
|\lceil v_i x\rfloor| \leq \psi.
$$
Thus one needs only to consider those $c$ satisfying $0<c \leq \lceil \psi  \rceil  + 1/2$
(this bound was given in \cite{SahG15} and it is easy to see actually it can be replaced by a
slightly tighter bound  $\lfloor \psi  \rfloor + 1/2$).
Therefore, if \eqref{e:xbound} holds, from \eqref{e:fastar}, we have
\beq
\label{e:fastar3}
f(\a^\star)=\min_{\xi_k,\xi_{k+1}  \in \Psi} f\left(\a \Big(\frac{\xi_k+\xi_{k+1}}{2}\Big)\right),
\eeq
where
\begin{align}
\label{e:phi}
& \Psi=\cup_{i=1}^n\Psi_i, \\
& \Psi_i=
\begin{cases}
\emptyset & v_i=0\\
\big\{\frac{c}{|v_i|}\big|0<c\leq \lceil \psi\rceil+\frac{1}{2}, c-\frac{1}{2}\in \mathbb{Z}\big\} & v_i\neq0
\end{cases}.
\label{e:phii}
\end{align}

The algorithm proposed in \cite{SahG15} for solving \eqref{e:oldop}
calculates the set $\Psi$ and sorts its  elements in increasing order and
then computes the right hand side of \eqref{e:fastar3},
and then compares it with $\min_{i} f(\pm e_i)$, which is equal to  $\min_{i} (1-\alpha v_i^2)$,
to get the solution.

By \eqref{e:oldop}, for $\a\in \mathbb{Z}^n$,
$$
f(\a)=\a^T\G\a=\sum_{i=1}^na_i^2-\alpha (\sum_{i=1}^na_iv_i)^2.
$$
According to \cite{SahG15}, since the discontinuity points of $f$ are sorted and at each step only one of the $a_i$'s change,
$\sum_{i=1}^na_i^2$ and $\alpha (\sum_{i=1}^na_iv_i)^2$ can be updated in constant time.
Therefore, $f(\a)$ can also be calculated in constant time.
Here we make a remark. Actually different $\Psi_i$ may have the same elements.
But they can be regarded as different quantities when $f(\a)$ is updated.
In order to remember which $a_i$ needs to be updated at each step,  a label is assigned to
every element of $\Psi$ to indicate which $\Psi_j$ it originally belonged to.
In the new algorithm proposed in Section \ref{sec:PA}, we will give more details about this idea.

Now we describe the complexity analysis given in \cite{SahG15} for the algorithm.
By \eqref{e:phii}, the number of elements of $\Psi_i$ is upper bounded by $\lceil \psi\rceil+1$,
so the number of elements of $\Psi$ is upper bounded by $n(\lceil \psi\rceil+1)$
(note  that if $v_i\neq 0$ for $1\leq i\leq n$, this is the exact number of elements).
From the above analysis, the complexity of the algorithm is determined by the sorting step,
which has the complexity of $\bigO(n\psi \log(n\psi))$,
where $\psi$ is defined in \eqref{e:psi}.

In the following, we derive the expected complexity of Algorithm SG
when $\h \sim \mathcal{N}(\0, \I )$ by following \cite{WenZMC15}.
Since $\h\sim \mathcal{N}(\0, \I )$,
$\norm{\h}_2^2\sim\chi^2(n)$.
Therefore, $\mathbb{E}[\norm{\h}_2^2]=n$.
Since $\sqrt{1+Px}$ is a concave function of $x$, by Jensen's Inequality,
\begin{align}
\label{e:JensenInequality}
\mathbb{E}\left[\psi\right]=\mathbb{E}\left[\sqrt{1+P\norm{\h}_2^2}\right]\leq \sqrt{1+nP}.
\end{align}
Thus,  the expected complexity of Algorithm SG is  $\bigO(n^{1.5}\log n)$.

\subsection{The Algorithm of Wen, Zhou, Mow and Chang}
\label{sec:EA2}
In this subsection, we review the algorithm of Wen et al given in \cite{WenZMC15},
an improvement of the earlier version given in \cite{WenZMC14}.
Its complexity is approximated by $\bigO(n (\log n+ \psi))$ based on the Gaussian heuristic,
where $\psi$ is defined in \eqref{e:psi}.
By \eqref{e:JensenInequality}, the expected complexity is approximately
$\bigO(n^{1.5})$ when $\h \sim \mathcal{N}(\0, \I )$.
For the sake of convenience, we will refer to the algorithm as Algorithm WZMC.

Again we want to simplify the matrix $\G$ in \eqref{e:CFP}. Define
\begin{equation}
\label{e:t}
\bst=\sqrt{\frac{P}{1 + P\norm{\h}_2^2}}\h = \sqrt{\alpha} \v.
\end{equation}
Then, \eqref{e:CFP} can be rewritten  as
\beq
\label{e:newop}
\min_{\a\in\mathbb{Z}^n\backslash\{\0\}}\a^T\G\a, \mbox{ where }
\G= \I - \bst \bst^T.
\eeq

Since $\h\neq\0$, $\bst\neq\0$.
Obviously, if $\a^\star$ is a solution to \eqref{e:newop}, then so is $-\a^\star$.
Thus, for simplicity, only the solution $\a^\star$ such that $\bst^T\a^\star\geq0$ was considered in \cite{WenZMC14} and \cite{WenZMC15}.
We also use this restriction throughout the rest of this paper.

In \cite{WenZMC15}, \eqref{e:newop} is first transformed to the standard form of the SVP \eqref{e:SVP}
by finding the Cholesky factor $\R$ of $\G$  (i.e., $\G=\R^T\R$)  based on the following theorem.
\begin{theorem}
\label{t:cholD}
The Cholesky factor $\R$ of $\G$ in \eqref{e:newop} is given by
\begin{eqnarray*}
r_{ij}=
\begin{cases}
\frac{g_i}{g_{i-1}}, & 1\leq j=i\leq n \cr
\frac{-t_it_j}{g_{i-1}g_i}, & 1\leq i<j\leq n \cr
 \end{cases}
,
\end{eqnarray*}
where $g_0=1$ and for $1\leq i\leq n$,
$g_i=\sqrt{1-\sum_{k=1}^{i}t^2_k}$.
\end{theorem}

Instead of forming the whole $\R$ explicitly, only the diagonal entries of $\R$ were calculated,
so it is easy to check that the complexity of this step is only $\bigO(n)$.

It was showed in \cite{RicSJ12} that if
\begin{align}
\label{e:tOrdered}
t_1\geq t_2\geq \ldots \geq t_n\geq 0,
\end{align}
then there exists a solution $\a^\star$ to \eqref{e:newop} satisfying
\begin{align}
\label{e:astarordered}
a^\star_1\geq a^\star_2\geq \ldots \geq a^\star_n\geq 0.
\end{align}
Given $\bst$,  if $t_i <0$ for some $i$, we can change it to $-t_i$
without changing anything else.
To have the order  \eqref{e:tOrdered},   we can permute the entries of $\bst $.
This  step costs $\bigO(n \log n)$.

To decrease the computational cost, the following $(n+1)-$dimensional vector $\p$
was introduced in \cite{WenZMC15}:
$$
p_{n+1}=0,\,\;p_i=p_{i+1}+t_ia_i,\,\; i=n, n-1, \ldots, 1.
$$
Define
$$
d_n=0, \; d_i=-\frac{1}{r_{ii}}\sum_{j=i+1}^nr_{ij}a_j, \quad i=n-1,\ldots, 1.
$$
Then, by Theorem \ref{t:cholD},
$$
d_i = \frac{t_i}{g_{i}}\sum_{j=i+1}^nt_{j}a_j=\frac{1}{g_{i}}t_ip_{i+1}, \quad i=n-1,\ldots, 1.
$$
Thus,
$$
\norm{\R\a}_2^2=\sum_{i=1}^n r_{ii}^2(a_i-d_i)^2 = \sum_{i=1}^n r_{ii}^2\Big(a_i-t_ip_{i+1}/g_i\Big)^2
$$
The Schnorr-Euchner search algorithm \cite{SchE94} was modified to search
the optimal solution satisfying \eqref{e:astarordered} within the ellipsoid:
$$
\sum_{i=1}^nr_{ii}^2 (a_i-t_ip_{i+1}/g_i)^2< f(\e_1) = 1-t_1^2.
$$
If no nonzero integer point is found,  $\e_1$ is the solution.
The cost of the search process was approximately $\bigO(n \psi)$  based on the Gaussian heuristic.
Thus, by the above analysis and \eqref{e:JensenInequality}, the expected complexity of Algorithm WZMC is approximated by
$\bigO(n^{1.5})$ when $\h\sim\mathcal{N}(\0, \I )$.

\section{New Algorithm}
\label{sec:PA}
In this section, we propose an algorithm with complexity of
$\bigO\big((n+\min\{\sqrt{n}, \varphi\} \varphi) \log (n+ \min\{\sqrt{n}, \varphi\} \varphi)\big)$,
where
\begin{align}
\label{e:varphi}
\varphi =\sqrt{1+P(\|\h\|^2-\max\limits_{1\leq i\leq n} h_i^2)}.
\end{align}
By \eqref{e:psi} and \eqref{e:JensenInequality}, the expected complexity is   $\bigO(n\log n)$ when $\h\sim\mathcal{N}(\0, \I )$.

Recall that Algorithm SG checks some discontinuity points of the objective function to find the optimal solution.
The main idea of our new algorithm is to reduce the number of discontinuity points to be checked.

In this following, we introduce a theorem, which can significantly reduce the number of discontinuity points
to be checked in finding the optimal one.
\begin{theorem}
\label{t:xbd1}
Suppose that $\bst$  satisfies
\beq \label{e:tineq}
t_1=\ldots =t_p > t_{p+1}\geq \ldots \geq t_{q} > t_{q+1} = \ldots = t_n =0,
\eeq
where both $p$ and $q$ can be 1 or $n$.
Then the solution $\a^\ast$ to \eqref{e:newop} satisfies either
$$
\a^\ast = \pm \e_k, \ \ 1\leq k \leq p,
$$
or
$$
\a^\star=\pm \lceil \bst x\rfloor
$$
for some $x\in \Rbb$ satisfying
\begin{align}
\label{e:xbd1}
\frac{1}{2t_2}\leq x\leq \mu,
\end{align}
where
\begin{align}
\label{e:alpha}
\mu=\min_{1\leq i\leq q } \left(\frac{1}{t_i}
\left\lfloor\frac{1}{\sqrt{i}}\sqrt{\frac{1-t_1^2}{1-\|\bst\|^2}}\right\rfloor+\frac{1}{2t_i}\right).
\end{align}

\end{theorem}
{\bf Proof.} Note that for $k\in \{1,\ldots,p\}$
$$
\min_{1\leq i\leq n} f(\pm \e_i) = \min_{1\leq i\leq n} (1-t_i^2) = 1-t_k^2 = f(\pm \e_k).
$$
It is possible that $\pm \e_k$ for $k=1,\ldots, p$ are  solutions to \eqref{e:newop}.
In the following proof we assume they are not.

By \eqref{e:t} and Theorem \ref{t:solpr}, there exists $x\in \Rbb$ such that  the solution can be written
as $\a^\star=\lceil \bst x\rfloor$.
Note that if $\lceil \bst x\rfloor$ is a solution, so is $-\lceil \bst x\rfloor$.
Thus we can just assume that $x$ here is positive.
Then by \eqref{e:tineq} we have
\beq \label{e:astarorder}
a^\ast_1 = \ldots=a^\ast_p \geq a^\ast_{p+1} \geq \ldots \geq a^\ast_q \geq a^\ast_{q+1} = \ldots =a^\ast_n =0.
\eeq
We must have $a^\ast_{2}  \geq 1$, otherwise $\a^\ast = \e_1$, contradicting with our assumption.
Thus,
$$
t_2x \geq  a^\star_2-\frac{1}{2}\geq\frac{1}{2}.
$$
Therefore, the first inequality in \eqref{e:xbd1} holds.

In the following, we show that the second inequality in \eqref{e:xbd1} holds.
Since $\e_1$ is not an optimal solution, $f(\a^\ast) < f(\e_1)$, i.e.,
$$
\|\a^\star\|^2-(\bst^T\a^\star)^2< 1-t_1^2.
$$
Therefore, by the Cauchy-Schwarz inequality,
\beq\label{e:aub}
\|\a^\star\|^2<\frac{1-t_1^2}{1-\|\bst\|^2}.
\eeq
By \eqref{e:astarorder}, for $ i=1,\ldots, q$,
\beq \label{e:alb}
\|\a^\star\|^2\geq  i  (a^\star_i)^2.
\eeq
Then, using the fact that  $\a^\ast = \lceil \bst x\rfloor$ and \eqref{e:aub} and \eqref{e:alb}, we have
$$
t_ix\leq a^\star_i+\frac{1}{2}\leq \left\lfloor\frac{1}{\sqrt{i}}\sqrt{\frac{1-t_1^2}{1-\|\bst\|^2}}\right\rfloor+\frac{1}{2}.
$$
Since the aforementioned equality holds for all $i=1,\ldots,q$,
the second inequality in \eqref{e:xbd1} holds, completing the proof.
$\Box$

Like Algorithm WZMC,  our new algorithm first performs a transformation on $\bst$ in \eqref{e:newop}
such that \eqref{e:tineq} holds,
costing $\bigO(n\log n)$.

 We define $\a(x)=\lceil \bst x\rfloor$ (cf.\  Section \ref{sec:EA1}).
Then for any $i=1,\ldots, q$, where $q$ is defined in \eqref{e:tineq}),
$a_i(x)=\lceil t_i x\rfloor$ is a piecewise constant function of $x$
and its discontinuity points are $x=\frac{c}{t_i}$ where $c-\frac{1}{2}\in \mathbb{Z}$.
To find the optimal discrete points, we need consider only a finite subset of those discrete points.
In fact,  by Theorem \ref{t:xbd1}, we  need to consider only those $x=c/t_i$, where $c$ satisfies
$$
\frac{t_i}{2t_2}\leq c \leq  t_i\mu, \ \  c-\frac{1}{2}\in \mathbb{Z}
$$
Thus, we define
\beq
\label{e:phibar}
\Phi=\cup_{i=1}^q {\Phi}_i,
\eeq
where
\beq
\label{e:phibari}
{\Phi}_i=
\Big\{\big (\frac{c}{t_i},i\big)  \, \Big| \, \frac{t_i}{2t_2} \leq c\leq  t_i\mu, \, c-\frac{1}{2}\in \mathbb{Z}\Big\}.
\eeq
Then the optimal discontinuity point  and its position in the vector $\a(x)$ must be in $\Phi$.
We sort the first elements of the members of ${\Phi}$ in increasing order,
then by \eqref{e:tOrdered}, only one entry of $\a$ increase $1$ for each element in ${\Phi}$
(note that if some of the entries of $\bst$ are the same, then the corresponding ${\Phi}_i$
have the same $x$, but we can regard them as different quantities to update $\a$ and the corresponding $f(\a(x))$).
By following \cite{SahG15}, we can compute $f(\a(x))$ for each $x$ by constant time.
Specifically, denote $T_1=\sum_{i=1}^na_i^2$ and $T_2=\sum_{i=1}^na_it_i$,  then $f(\a(x))=T_1-T_2^2$.
We start from $T_1=T_2=0$ and $\a=\0$, and for each $(x,i)\in {\Phi}$, we
update $\a$ by setting $a_i=a_i+1$, then we update $T_1$ by setting $T_1=T_1+2a_i-1$,
update $T_2$ by setting $T_2=T_2+t_i$, and update $f$ by setting $f=T_1-T_2^2$.
During the enumeration process, we only keep the $\a$ which minimizes $f$ and the corresponding $f$.
If $f<1-t_1^2$, then the final $\a$ is $\a^\star$, otherwise, $\a^\star=\e_1$.

By the above analysis, the algorithm can be summarized in Algorithm \ref{alg:PA1}.

\begin{algorithm}[htb]
\caption{New Algorithm}
\label{alg:PA1}
\renewcommand{\algorithmicrequire}{ \textbf{Input:}}      
\renewcommand{\algorithmicensure}{ \textbf{Output:}}     
\begin{algorithmic}[1]
\REQUIRE ~~Channel vector $\h$ and transmission power $P$
\ENSURE ~~$\a^\star$
\renewcommand{\algorithmicrequire}{ \textbf{Initialization:}} 
\REQUIRE ~~\\
\STATE calculate $\bst$ by \eqref{e:t}
\STATE perform a signed permutation on $\bst$ such that the new $\bst$
satisfies \eqref{e:tineq}
\STATE calculate $\mu$ by \eqref{e:alpha}
\STATE let $\Phi=\emptyset$, $f_{min}=1-t_1^2$, $\a^\star=\e_1$

\renewcommand{\algorithmicrequire}{ \textbf{Phase 1:}} 
\REQUIRE ~~\\
\FOR{all $i \in \{1,...,q\}$}
    \FOR{all $c-1/2 \in \mathbb{Z}$ such that  $t_i/(2t_2) \leq c\leq  t_i\mu$ }
        \STATE calculate $x=c/t_i$
        \STATE $\Phi=\Phi\cup{(x,i)}$
    \ENDFOR
\ENDFOR

\renewcommand{\algorithmicrequire}{ \textbf{Phase 2:}} 
\REQUIRE ~~\\
\STATE \label{a:sortphi} sort $\Phi$ by the first element of the members in an increasing order
\STATE set $T_1=0$, $T_2=0$ and $\a=\0$.
\FOR{every  $(x,i)\in\Phi$}
    \STATE $a_i=a_i+t_i$
    \STATE $T_1=T_1+2a_i-1$
    \STATE $T_2=T_2+t_i$
    \STATE $f=T_1-T_2^2$
    \IF{$f<f_{min}$}
        \STATE set $\a^\star=\a$
        \STATE set $f_{min}=f$
    \ENDIF
\ENDFOR
\RETURN sign permuted $\a^\star$

\end{algorithmic}
\end{algorithm}

Before analyzing the complexity of Algorithm \ref{alg:PA1},
we look into the number of discontinuous points needed to be checked by  Algorithm \ref{alg:PA1} and Algorithm SG.
By \eqref{e:alpha}, \eqref{e:t} and \eqref{e:varphi}, for $i \in \{1,\ldots,q\}$,
\begin{align*}
 t_i\mu -\frac{1}{2}
&\leq  \left\lfloor\frac{1}{\sqrt{i}}\sqrt{\frac{1-t_1^2}{1-\|\bst\|^2}}\right\rfloor \\
&= \left\lfloor\frac{1}{\sqrt{i}}\sqrt{1+P(\|\h\|^2- \max_{j} h_j^2)}\right\rfloor
=  \left\lfloor\frac{1}{\sqrt{i}} \varphi \right\rfloor, \nonumber
\end{align*}
where the ``max'' is involved because $t_1^2$ is the largest among all $t_i^2$ after the permutation of $\bst$ (see \eqref{e:tOrdered}).
Thus, by \eqref{e:phibari},
\begin{align}
\label{e:|phiibar|}
|{\Phi}_i| \leq  \left\lfloor \varphi/ \sqrt{i} \right\rfloor + 1.
\end{align}

By \eqref{e:phii}, for $i \in \{1,\ldots,q\}$,
\beq
\label{e:|phii|}
|\Psi_i| = \lceil \psi \rceil +1,
\eeq
where $\psi$ is defined in \eqref{e:psi}.

Thus, from \eqref{e:|phiibar|}, \eqref{e:|phii|}, \eqref{e:psi} and \eqref{e:varphi},  it follows that
$$
\frac{|{\Phi}_i|}{|\Psi_i|}\leq   \frac{\left\lfloor \varphi/ \sqrt{i} \right\rfloor + 1}{\lceil \psi \rceil + 1}<1,  \ \
i =1, \ldots, q
$$
Note that $\varphi$ can be arbitrarily smaller than $\psi$ (see \eqref{e:psi} and \eqref{e:varphi}).
Also when $i$ is big enough, $\lfloor \varphi/ \sqrt{i} \rfloor =0$.
Thus  the new algorithm can significantly reduce the number of
discontinuity points to be checked.

Now we study the complexity of Algorithm \ref{alg:PA1}.
By \eqref{e:|phiibar|}, when $i > \lfloor \varphi^2 \rfloor$, $ |\Phi_i | =0.$
Then, with $k=\min\{q, \lfloor \varphi^2 \rfloor\}$, by \eqref{e:|phiibar|}, we have
\begin{align*}
|\Phi |& = \sum_{i=1}^q |\Phi_i| \leq \varphi\sum_{i=1}^{k } \frac{1}{\sqrt{i}} + q
 \leq  \varphi\sum_{i=1}^{k}\int_{i-1}^i \frac{1}{\sqrt{x}}dx + q \\
& =  2\sqrt{k}\varphi + q \leq 2  \min\{\sqrt{n}, \varphi\} \varphi+n .
\end{align*}

Thus the complexity of line \ref{a:sortphi} of Algorithm 1 is
$\bigO((n+\min\{\sqrt{n}, \varphi\} \varphi) \log (n+ \min\{\sqrt{n}, \varphi\} \varphi)$.
It is easy to see it is actually the complexity of the whole algorithm.
Then it follows from  \eqref{e:psi}, \eqref{e:varphi} and \eqref{e:JensenInequality}
that the expected complexity of Algorithm 1 is $\bigO(n\log n)$ when $\h\sim\mathcal{N}(\0, \I )$.

\section{Numerical simulations}
\label{sec:sim}
In this section, we present numerical results to compare the efficiency of our proposed method Algorithm \ref{alg:PA1} (denoted by "Proposed")
with those in \cite{SahG15} (denoted by "SG") and \cite{WenZMC15} (denoted by "WZMC").
We do not compare Algorithm \ref{alg:PA1} with the  branch-and-bound algorithm in \cite{RicSJ12} since
numerical tests in \cite{WenZMC15} show that the algorithm in \cite{SahG14} is faster,
while the algorithm in \cite{SahG15} is an improved version of that in \cite{SahG14}.
All the numerical tests  were done by \textsc{Matlab} 2010a on a laptop with Intel(R) Core(TM) i5-5300U CPU@ 2.30GHz.

\begin{figure}[!htbp]
\centering
\includegraphics[width=3.2in]{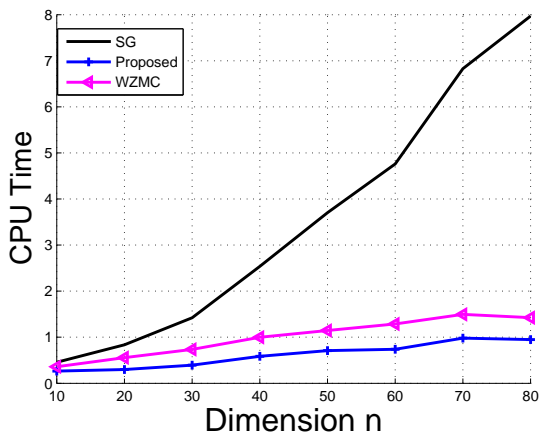}
\caption{Total CPU time   versus $\n$ for $P=1$} \label{fig:P=1}
\end{figure}

We set the dimension $n$ of $\h$ being $10, 20, \ldots, 80$ and $\h \sim \mathcal{N}\left(\0,\I\right)$.
For each given $n$ and $P$, we randomly generate $1000$ realizations of $\h$  and apply the three algorithms to solve \eqref{e:CFP}.
Figure \ref{fig:P=1} shows the {\em total} CPU time for $P=1$.

From Figure \ref{fig:P=1}, we can see that the total CPU time for the proposed method and WZMC are very close,
So for comparisons we also give Table \ref{tb:P=50} to show the {\em total} CPU time for $P=50$.

\begin{table}[htbp!]
\caption{Total CPU time in seconds  versus $\n$ for $P=50$}
\centering
\begin{tabular}{|c|c|c|c|c|c|c|c|c|c|c|}
 \hline
\backslashbox{}{$n$}& $10$ & $20$ & $30$& $40$& $50$ & $60$& $70$& $80$\\ \hline
SG  & 2.77 & 6.28 & 13.3 & 23.5 & 41.4 & 62.7&93.7&130 \\ \hline
Proposed &0.96 & 1.40 & 2.06 & 2.60 & 3.42 & 4.15 & 5.14&6.07 \\ \hline
WZMC  & 1.72 & 2.31 & 2.28 & 2.49 & 2.62 & 2.77&3.14&3.42 \\ \hline
\end{tabular}
\label{tb:P=50}
\end{table}

From Figure \ref{fig:P=1} and Table \ref{tb:P=50}, we can see that our proposed algorithm is much faster than SG,
and it is also faster than WZMC if both $n$ and $P$ are not very large
which means the new algorithm and WZMC have advantages in different settings, so both of them are useful.
Algorithm \ref{alg:PA1} has another advantage, i.e., its complexity can be rigorously analyzed,
while the complexity for WZMC was based on Gaussian heuristic.

\section{Conclusions}
\label{sec:Con}
In this paper, we proposed an algorithm with the expected complexity of $\bigO(n\log n)$ for a shortest vector problem
arising in compute-and-forward network coding design.
The complexity is lower than  $\bigO(n^{1.5}\log n)$, the expected complexity of the latest algorithm in \cite{SahG15}
and $\bigO(n^{1.5})$, the approximate expected complexity of  the algorithm proposed in  \cite{WenZMC15}.
Simulation results showed that the new  algorithm is much faster than that given in \cite{SahG15}.

\bibliographystyle{IEEEtran}
\bibliography{ref}

\end{document}